# Cases of EFL Secondary Students' Prompt Engineering Pathways to Complete a Writing Task with ChatGPT

David James Woo [a, *], Kai Guo [b] and Hengky Susanto [c]

[a] Precious Blood Secondary School, 338 San Ha Street, Chai Wan, Hong Kong, China

[b] Faculty of Education, The University of Hong Kong, Hong Kong, China

[c] Education University of Hong Kong, Hong Kong, China

[*] Corresponding author

- Postal address: Precious Blood Secondary School, 338 San Ha Street, Chai Wan, Hong Kong, China

- Email address: net_david@pbss.hk

- Phone: +852 2570 4172

## Author Bio

***David James Woo*** is a secondary school teacher. His research interests are in artificial intelligence, natural language processing, digital literacy, and educational technology innovations. ORCID: https://orcid.org/0000-0003-4417-3686

***Kai Guo*** is a Ph.D. candidate in the Faculty of Education at the University of Hong Kong. His research focuses on second language writing, technology-enhanced learning, and artificial intelligence in education. His recent publications have appeared in international peer-reviewed journals such as *Computers & Education*, *Interactive Learning Environments*, and *Journal of Educational Computing Research*. ORCID: https://orcid.org/0000-0001-9699-7527

***Hengky Susanto*** received his BS, MS and PhD degree in computer science from the University of Massachusetts system. He was a post-doctoral research fellow at University of Massachusetts Lowell and Hong Kong University of Science and Technology. He was also senior researcher at Huawei Future Network Theory Lab. Currently, he is a principal researcher in a startup mode research laboratory and a lecturer at Education University of Hong Kong. His research interests include applied AI (computer vision and NLP) to solve complex social problems, smart city, and computer networking (e.g., datacenter network, congestion control, etc.).

## Conflict of Interest Statement

We have no conflicts of interest to disclose. This research received no specific grant from any funding agency in the public, commercial, or not-for-profit sectors.

## Acknowledgements

The authors thank Chi Ho Yeung and Lancer for their data analysis suggestions.

# Cases of EFL Secondary Students' Prompt Engineering Pathways to Complete a Writing Task with ChatGPT

## Abstract

ChatGPT is a state-of-the-art (SOTA) chatbot. Although it has potential to support English as a foreign language (EFL) students' writing, to effectively collaborate with it, a student must learn to engineer prompts, that is, the skill of crafting appropriate instructions so that ChatGPT produces desired outputs. However, writing an appropriate prompt for ChatGPT is not straightforward for non-technical users who suffer a trial-and-error process. This paper examines the content of EFL students' ChatGPT prompts when completing a writing task and explores patterns in the quality and quantity of the prompts. The data come from iPad screen recordings of secondary school EFL students who used ChatGPT and other SOTA chatbots for the first time to complete the same writing task. The paper presents a case study of four distinct pathways that illustrate the trial-and-error process and show different combinations of prompt content and quantity. The cases contribute evidence for the need to provide prompt engineering education in the context of the EFL writing classroom, if students are to move beyond an individual trial-and-error process, learning a greater variety of prompt content and more sophisticated prompts to support their writing.

## 1. Introduction

ChatGPT's incredible popularity indicates many people's desire to transform their world of education, work and leisure through chatbots. Previously, chatbots followed a rule-based design with limited capabilities to respond accurately to user queries, especially with unfamiliar inputs. In contrast, state-of-the-art (SOTA) generative artificial intelligence (AI) chatbots like

ChatGPT rely on neural network language models that have been trained on a large corpus of data and can improve the accuracy of their responses (Caldarini et al., 2022). As ChatGPT has become the standard of performance for SOTA chatbots, ChatGPT is becoming a catch-all term for SOTA chatbots.

To effectively and ethically use SOTA chatbots in everyday life requires AI literacy (Long & Magerko, 2020) and people must learn the knowledge, skills and attitudes necessary to interact with these chatbots (Vuorikari et al., 2022). To access a SOTA chatbot's capabilities to perform strongly in human language tasks and even generate texts that are indistinguishable from human-written texts (Brown et al., 2020), people must learn to prompt the chatbot, that is, to deliver a set of instructions to guide the chatbot's text generation for a specific task (White et al., 2023).

Prompt engineering refers to the process of crafting an appropriate prompt so that a chatbot produces a desired output (Liu et al., 2021). Prompt engineering is a skill that requires expertise and practice to learn (OpppenLaender et al., 2023). In spite of humans being the most effective prompt engineers, the challenge is neither prompt-engineering instruction nor effective prompt-engineering practice has become widespread (Zhou et al., 2023). As a result, for the non-technical user, writing an appropriate prompt for a SOTA chatbot like ChatGPT is not straightforward and non-technical users suffer a trial-and-error process (Dang et al., 2022).

In this paper, we report cases of non-technical users' trial-and-error prompt engineering. The cases were selected from the context of English as a foreign language (EFL) students in a secondary school class, completing a timed-writing task and using SOTA chatbots for the first time, although the students previously used other types of generative AI to complete writing tasks. Each case illustrates a distinct prompt engineering pathway, that is, a different combination of prompt content and total number of prompts to complete the same writing task. Comparing the cases, we find some pathways appear more sophisticated than others. By understanding the similarities and differences in cases, educators can develop principles for prompt engineering education and cater to individual, EFL students' learning needs, facilitating these students development of effective strategies for completing writing tasks with ChatGPT and other SOTA chatbots.

## 2. Literature Review

### 2.1. Using ChatGPT to Support EFL Writing

Writing is a crucial skill for EFL students, but they often encounter challenges when writing in a language that is not their native one. These challenges include struggling with grammar, vocabulary, syntax (De Wilde, 2023; Vasylets & Marín, 2021), a lack of confidence in their writing abilities (Sun & Wang, 2020; Zotzmann & Sheldrake, 2021), and encountering difficulties in generating ideas during the writing process (Crossley et al., 2016; Hayes & Flower, 2016). Previous research has investigated different pedagogical approaches to enhance the writing skills of EFL students, and collaborative writing has been extensively implemented, exhibiting favorable effects on students' writing performance (for a meta-analysis, see Elabdali, 2021). Nevertheless, in practice, it can be challenging for students to identify an ideal writing partner, leading some scholars to propose the utilization of chatbots as a learning companion to assist students in their writing (e.g., Author, 2022; Su et al., 2023).

A chatbot is a conversational user interface that enables human users to engage in meaningful verbal or text-based exchanges with a computer program (Kim et al., 2022). In this exchange or *turn*-taking, a user provides input and a chatbot responds, mimicking human dialogue. For EFL students in a writing classroom, chatbots may have capability to provide immediate and personalized feedback, fostering a non-judgmental and supportive learning environment (Bibauw et al., 2022). Furthermore, chatbots can generate ideas, and importantly genre-specific ideas as Author's (2022, 2023a) Argumate chatbot could provide ideas that refute an opposing view in an argumentative essay. Similarly, Zhang et al.'s (2023) chatbot could provide EFL students with self-regulated training on logical fallacies in argumentative writing.

  Since its launch in November 2022, ChatGPT, a chatbot utilizing SOTA language models, has sparked extensive discussions among education researchers. This is not least because of its impressive technical features, such as memory to recall previous interactions and ability to interpret the nuances of human language, both of which enable ChatGPT to generate coherent and contextually appropriate responses. Particularly, some researchers are exploring how ChatGPT can support students' writing. For example, Su et al. (2023) explored how ChatGPT can assist at different stages of a student's writing process, including outline preparation, content revision, proofreading, and post-writing reflection. Yan (2023) investigated EFL students' exposure to ChatGPT in writing classrooms and their resulting behaviors and reflections. The results highlighted the effectiveness of ChatGPT as a tool for EFL writing instruction, showcasing its potential to optimize the efficiency of writing composition through its automated workflow.

  While ChatGPT and other SOTA chatbots can address some problems EFL students face when writing, our view is ChatGPT should not completely replace EFL students' writing. Instead, we frame EFL students' use of ChatGPT to complete a writing task as a "machine-in-the-loop", where ChatGPT and students collaborate to produce written text (Clark et al., 2018; Yang et al., 2022). Figure 1 illustrates a student completing a writing task with a machine-in-the-loop. First, a student begins a turn by prompting ChatGPT to generate text output. After the prompting, the student critically evaluates the generated output and integrates it into their writing as necessary. The student repeats these turns until the writing task is complete and these turns compose a pathway. Besides, the student may have access to SOTA chatbots besides ChatGPT. Importantly, "machine-in-the-loop" writing may foster synergy between the strengths of chatbots and the student, resulting in higher-quality writing, but this depends not least on whether a student strategically develops prompts that guide ChatGPT to generate relevant output text (Author, 2023b).

**Figure 1**

*A writing task with a "machine-in-the-loop"*

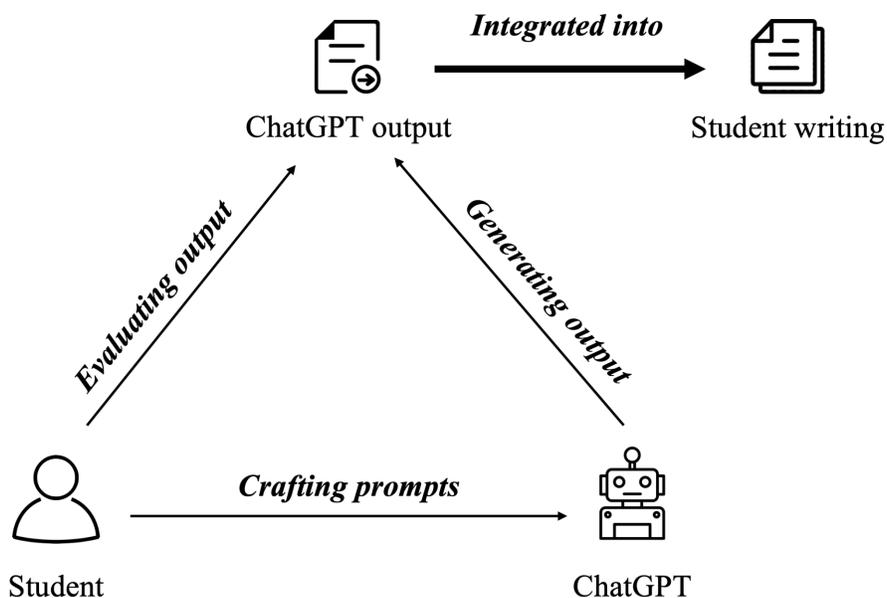

## 2.2. Prompt Engineering

Although prompts can be designed for a variety of human language tasks that SOTA chatbots can perform, Saravia (2022) proposed that a prompt can combine two main components. The first component is *context*, which helps ground a chatbot's output in a knowledge base. Context has often included input data or output examples so that, for instance, *one-shot* learning has referred to a chatbot's ability to learn concepts to perform a task from encountering one example (Lampinen & McClelland, 2018); and *few-shot* learning has referred to a chatbot's ability to generalize from a few examples (Parnami & Lee, 2022).

The second component is *natural language*, that is, writing an explicit prompt in a human's native language because of chatbot's increasing capability to understand abstract task descriptions and human concepts (Reynolds & McDonell, 2021). An essential type of natural language prompting is direct task specification or natural language instruction, which refers to imperative commands that specify the type of output desired, for instance, "generate," "list," "write," etc. (Cohere, 2023). This approach allows users to communicate with a chatbot in a way that is intuitive and more closely resembles human-to-human interaction. Chatbots have increasingly been trained to respond capably to natural language instruction for up to 175 types of language tasks (Wang et al., 2023). Instructing a chatbot in natural language without any task specific exemplars can be considered a *zero-shot* learning task for the chatbot.

Because of natural language prompting, users can supply additional context to a prompt, beyond one-shot and few-shot, that may unlock a SOTA chatbot's novel capabilities (Reynolds & McDonell, 2021). One such capability is chain-of-thought (COT) reasoning, that is, breaking down a problem into steps before delivering a verdict. Kojima et al. (2022) found adding "Let's think step-by-step" at the end of a prompt could unlock chatbots' COT reasoning capabilities across diverse tasks without few-shot learning. For ChatGPT specifically, researchers recommend strategies such as prompting ChatGPT to take on a specific perspective or role (Akin, 2023). Besides, because natural language can effectively prompt a SOTA chatbot, an important principle for effective prompting is choosing words carefully (Atlas, 2023). Ultimately, natural language instruction and context can be ordered differently and layered depending on the task and desired output.

In conclusion, recent research has suggested various principles and strategies for effectively engineering prompts. However, in the context of EFL students who are completing a writing task with ChatGPT, we have no evidence of the quality, quantity and sequence of these students' prompts to generate desired output. To take these students beyond an individual's trial-and-error approach, there is a need to examine how these students can engineer prompt pathways.

**2.3. Research Questions**

1) What types of content do EFL students use to prompt chatbots during a writing task?
2) What patterns emerge in the prompt engineering pathways of EFL students when completing a writing task using chatbots?

## 3. Research Design and Methodology

**3.1. Case Study**

We use a qualitative, case study approach (Creswell, 2013) that provides context and a thick description for EFL students' prompt engineering pathways. Since this study conceptualizes prompt engineering pathways as turn-taking within a machine-in-the-loop framework, the case unit of analysis is each exchange or turn where a student prompts a chatbot. A case focuses on one student taking turns to prompt a chatbot so as to complete a writing task. In terms of outcomes, the study is interested in the number of prompts, the types of prompts and the sequencing of prompts as well as the quantity and quality of each prompt type involved in a student's prompt engineering pathway.

We use multiple cases to add more data to answer the research questions. We selected four cases that are drawn from the same educational context, but illustrate differences in prompt engineering pathways.

**3.2. Procedures**

A purposive, convenience sampling technique (Cohen et al., 2017) was used for the selection of EFL students. The first author recruited four EFL students from a form one class in a Hong Kong secondary school, which generally recruits students at the 44th to 55th percentile of academic achievement in its school district. Within the school, the four students were the highest-achieving students in their form for the English language subject area. The four students have had previous experience completing writing tasks in their English lessons using open-source text generators and text-to-image generators but had never used SOTA chatbots to complete a writing task.

The students were instructed to compose a blog entry, a text type they had learned and practiced prior to the study, using their own words and words from chatbots (see Appendix 1). They were not given explicit instruction on the capabilities of the chatbots, how chatbots are different from other generative AI and how to prompt a chatbot. Each student was given an iPad. The students wrote their blog entries on Google Docs. As for chatbots, the students used the POE app, which at the time of the study comprised a suite of proprietary chatbots including ChatGPT, Sage and Dragonfly. Students completed the task in a classroom at an after-school lesson on April 21, 2023. The students had at most 45 minutes to complete the task and all students completed the task. The first author taught the lesson.

**3.3. Data Collection**

Our central form of data was screen recordings of students' iPads as the students completed the writing task. For data triangulation, we collected students' completed writing tasks on Google Docs (see Appendix 2).

**3.4. Data Analysis**

To address the first research question, we performed a content analysis (Joffe & Yardley, 2004) of students' prompts as evidenced in screen recordings. For data preparation, the second author viewed screen recordings, transcribed each student's prompt verbatim and sequenced the prompt in a Google Sheet. The second author also noted which chatbot a student used for each prompt. The first, second and third authors open-coded (Saldana, 2012) each prompt, with each code representing a different type of prompt content. We applied deductive codes from our literature review as well as inductive codes. Besides, a prompt could receive more than one code. For more valid and reliable coding, the first author developed a codebook (DeCuir-Gunby et al., 2010) and the first and second authors performed an inter-rater agreement test (Campbell et al., 2013). After discussion, the authors reached one-hundred percent agreement (see Appendix 3) on the application of codes to prompts. From this coding, we report overall and for each case the number of prompts, the types of prompts and the quantity and quality of each prompt type and the sequencing of prompts.

To answer the second research question, we visualized the coded data so as to facilitate pattern identification. We visualized each case's data as a network graph, which is commonly used to represent the relationships between different objects. The network graph consists of *nodes* and *edges*. The nodes in our network graphs are circles and symbolize distinct types of prompt content. The edges represent the connections between two nodes. The direction of an edge is denoted by an arrow, which specifies the sequence in which the prompt content types are performed by a student. A solid line between two nodes indicates a student has prompted the same chatbot as the previous turn and a dotted line indicates a student has prompted a different chatbot from the previous turn.

Examining each case's network graph, we identified intra-case patterns, that is, continuity and change in prompt types and chatbots used. For each case, we report these patterns and supplement these with word and topic patterns, that is, continuity and change in the words and topics of prompts, identified from a careful reading of each case's prompts.

For cross-case analysis, we applied pattern-matching (Yin, 2003) which refers to identifying dependent variables and patterns which lead to similar outcomes in the dependent variables across cases. Share patterns between cases can be considered a theoretical replication.

## 4. Findings

### 4.1. Types of Prompts

In total, the four students took 38 turns with chatbots to complete their writing task. However, we discarded prompts from two of those turns from our analysis because the first and second authors had identified these instances as students accidentally clicking prompts automatically generated and suggested by chatbots.

From 36 coded turns, we found different types of prompt content. In Table 1 different types of prompt content are listed in alphabetical order by code name. We found six distinct types of prompt content (see code numbers one, four, five, seven, eight and nine). In addition, we found more sophisticated types of prompt content (code numbers two, three and six) which are developed from natural language instruction in combination with other distinct types of prompt content.

**Table 1**

*Coding scheme*

| Code No. | Code | Definition | Example | No. of instances |
|---|---|---|---|---|
| 1 | AC | Auto-complete: a one-shot prompt of an input data exemplar without natural language instruction and without a question. The intent is for the AI to learn to generate output for the input data exemplar by autocomplete, that is, extension. | *I read a book this morning at school..* (Student C) | 8 |
| 2 | AC + NLI | One-shot: an auto-complete prompt with natural language instruction and an input data exemplar by which the AI learns to generate output | *I read a book this morning at school. Continue the story and answer this question: with whom did you do it with?* (Student C) | 2 |
| 3 | AC + NLI + Q | Natural language instruction with a question and an input data exemplar | *I was breathing while waiting to go up to the classroom. Continue the story and answer this question : why did you do it?* (Student D) | 8 |
| 4 | GRE | Greeting: a prompt that acknowledges the chatbot as if it were a human but that does not appear to facilitate completion of the writing task | *hello* (Student B) | 2 |
| 5 | NLI | Natural language instruction: explicit instruction in natural language with an imperative verb and ZERO shot, that is, without an input data exemplar | *Write a blog entry about my morning* (Student B) | 7 |
| 6 | NLI + Q | Natural language instruction with a question | *Write a blow (sic) entry about why going to school is a good thing and how it feels like.* (Student B) | 1 |
| 7 | Q | Question: a prompt formulated with either a direct or indirect question, that is, a statement beginning with a question word | *What did you do this morning* (Student B) | 2 |
| 8 | SE | Search engine: a prompt for information, grounded in the real world, factual; not formulated as a question | *Some famous place in turkey* (Student A) | 4 |
| 9 | TMM | Tell me more: a prompt automatically generated and suggested to a user by a chatbot; all TMM prompts rely on a previous turn as input data for the AI | *Tell me more.* (Student D) | 2 |

The most popular prompt types were auto-complete prompts (code no. one) and one-shot prompts with natural language instruction and a question (code no. three) as eight prompts were coded for each. The least popular prompt type was natural language instruction with a question (code no. six) as one prompt was coded for that.

**4.2. Prompt Engineering Pathways**

Figure 2 is a line chart to visualize the distribution of prompts by student. Each line represents a student's prompt engineering pathway. Each line is mapped to an X-axis, which shows the sequence of each student's prompts from the first to the last. In this chart, we observe that Student A took the fewest turns (n = 5) to complete the writing task and Student B (n = 12) took the most. Each line is also mapped to a Y-axis, which shows code numbers. We observe no student had used all prompt types and each student had used a different combination of prompt types to complete the writing task. We elaborate each student's pathway in the following sections, with code names given in parentheses.

**Figure 2**

*Four students' prompt engineering pathways*

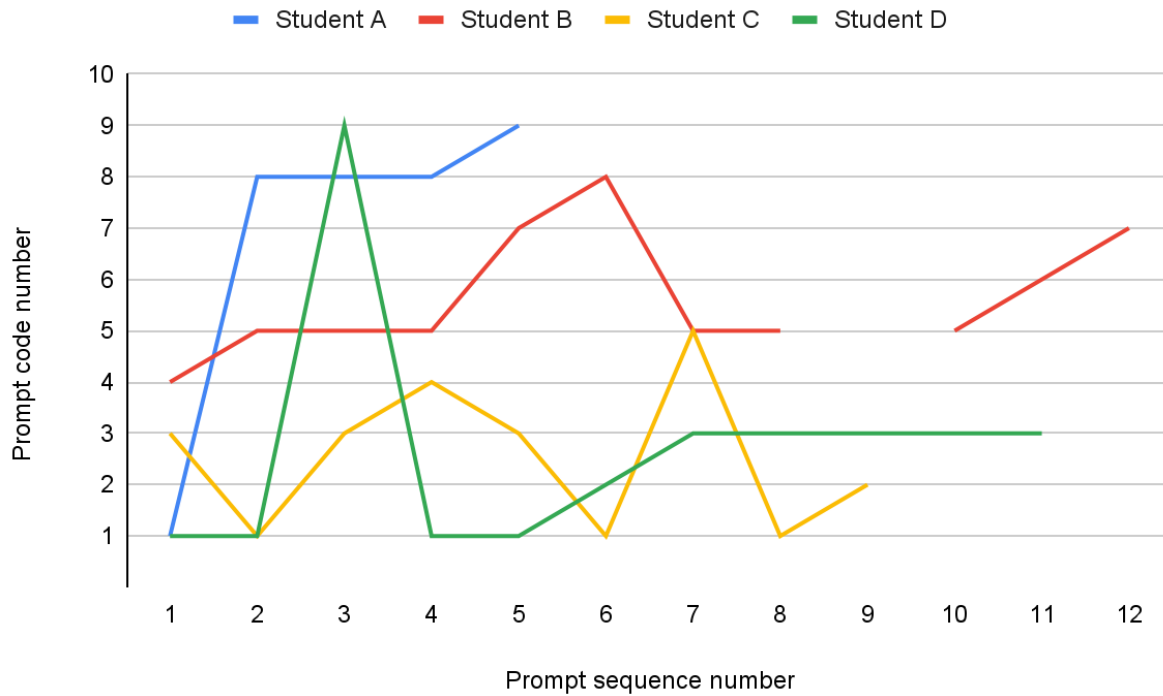

*Note*: Student B's ninth prompt and Student C's tenth were omitted from the analysis as the first and second author observed these students had accidentally prompted their chatbots on these turns.

### *4.2.1. Student A's Prompt Engineering Pathway*

Student A was writing a blog entry about Turkey and its attractions. Student A used only the Sage chatbot.

Figure 3 is a network graph for Student A's prompt engineering pathway. Student A took five turns to complete the writing task and wrote three types of prompts. For the first turn, the student wrote a prompt, "I woke up and," without natural language instruction (AC), intending for the chatbot to autocomplete the prompt.

**Figure 3**

*Network graph of Student A's prompt engineering pathway*

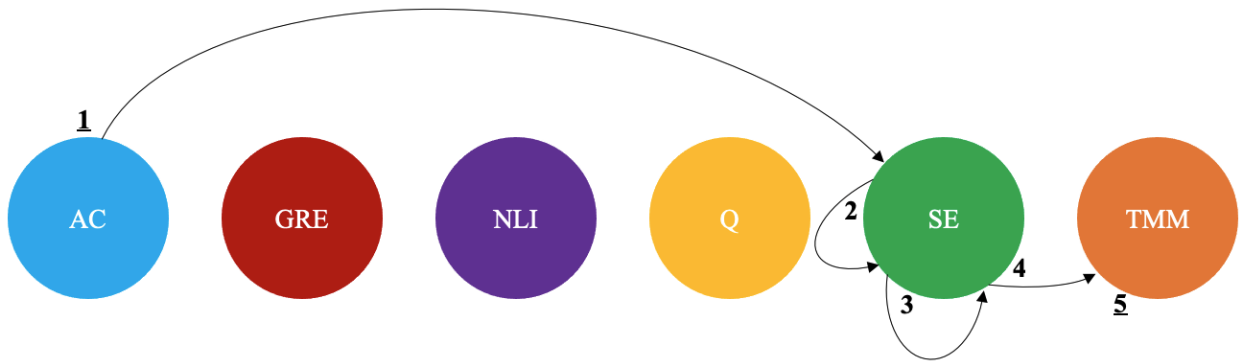

For turns two through four, Student A wrote prompts to elicit factual information from the chatbot (SE), treating the prompt as if it were a Google search. The words in this series of prompts are completely different from the words of the first turn's prompt. However, the words in this series of prompts show much continuity, as Student A carefully and systematically changed words from prompt two onwards to inquire about topics in Turkey:

- Some famous places in turkey (turn two)
- Famous drinks in Turkey (turn three)
- Some fun activities to do in turkey on vacation (turn four)

In turn five Student A selected a prompt automatically suggested by the chatbot (TMM), "Tell me more." This appears to be the first and only time that Student A makes use of the chatbot's chat memory.

*4.2.2. Student B's Prompt Engineering Pathway*

Student B was writing a blog entry about her morning. Student B used only the Sage chatbot.

STUDENTS' PROMPT ENGINEERING PATHWAYS                                                                 17Figure 4 is a network graph for Student B's prompt engineering pathway. Student B took 12 turns to complete her writing task and wrote five types of prompts. For the first turn, Student B wrote a greeting to the chatbot (GRE), "hello."

**Figure 4**

*Network graph of Student B's prompt engineering pathway*

*Note*: Student B's ninth prompt was omitted from the analysis as the first and second author observed this student had accidentally prompted their chatbot on this turn.

For turns two through four, Student B wrote explicit instructions for the chatbot, beginning with imperative verbs and without input data exemplars (NLI). The words in this series of prompts are completely different from the word of the first turn's GRE prompt. However, the words in this series of prompts show much continuity, as Student B carefully and systematically changed words from prompt two onwards to instruct the chatbot:

- Write a blog entry about a day going to secondary school (turn two)
- Write a short blog entry about my day at school (turn three)

- Write a blow (sic) entry about. My day to school (turn four)

For the fifth turn, Student B wrote a question (Q), "What did you do this morning," which shows discontinuity from the previous prompts not only in prompt type but words. For the sixth turn, Student B wrote a prompt, "Ideas to do in the morning before school," which elicits information from the chatbot (SE). Although the prompt type is different from previous prompts, the words in the prompt show continuity with words in prompts two through five.

For turns seven, eight and 10, Student B reverted to writing natural language instruction (NLI). This series of prompts shows continuity in the instructions, prompts seven and eight shows continuity in the words from prompts five and six:

- Write a blog entry about my morning (turn seven)
- Write a blog entry about morning (turn eight)
- Write a blow (sic) entry about meeting friends is happy (turn 10)

For turn 11, Student B combined types of prompts for the first and only time, writing natural language instruction with two questions (NLI + Q) for the chatbot to address, "Write a blow (sic) entry about why going to school is a good thing and how it feels like." The words for the NLI show continuity with Student B's previous NLI prompts, and the topical word school shows continuity with prompts two, three, four and six.

Finally, for turn 12, student B wrote a question (Q), "Why learning is fun," which shows discontinuity with words from previous prompts.

### *4.2.3. Student C's Prompt Engineering Pathway*

Student C wrote a blog entry about her morning. Student C used the Sage, ChatGPT and Dragonfly chatbots.

Figure 5 is a network graph for Student C's prompt engineering pathway. Student C took nine turns to complete her writing task and wrote five types of prompts.

**Figure 5**

*Network graph of Student C's prompt engineering pathway*

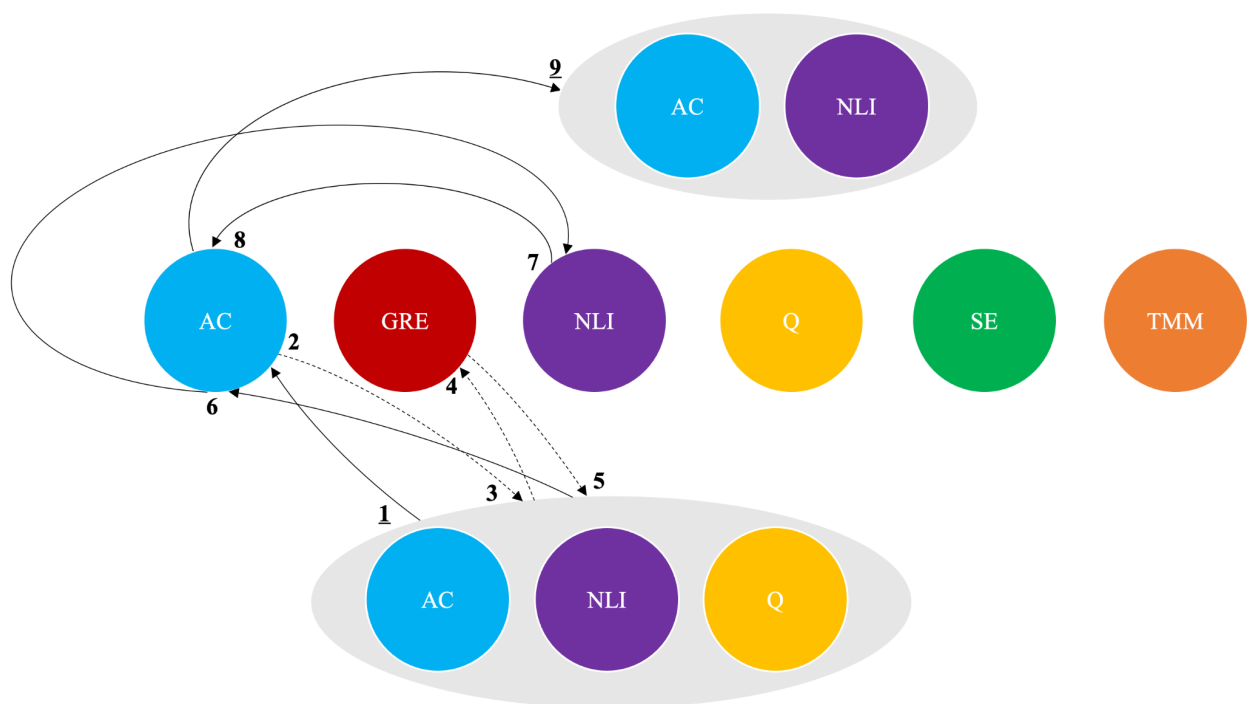

*Note*: Student C's tenth prompt was omitted from the analysis as the first and second author observed Student C had accidentally prompted their chatbot on this turn.

Student C began with Sage chatbot. For turn one, Student C wrote natural language instruction with a one-shot exemplar and a question (AC + NLI + Q), "I read a book this morning at school. Continue the story and answer this question : with whom did you do it with?"

For turn two, Student C wrote a prompt without natural language instruction (AC), with an intent for the chatbot to autocomplete the prompt. In fact, the prompt is the one-shot exemplar from the first prompt, "I read a book this morning at school.." However, the second prompt is stripped of the natural language instruction and the question from the first prompt.

For turn three, Student C used ChatGPT. Student C reverted to the exact words of the first turn's AC + NLI + Q prompt.

For turn four, Student C reverted to Sage chatbot. Student C wrote a greeting to the chatbot (GRE), "Sage hi." The words in this prompt show discontinuity with previous prompts and this is the only instance in the study where a student greeted a chatbot by name.

From turn five, Student C used Dragonfly chatbot. Student C reverted to the exact words of the first and third turn's AC + NLI + Q prompt.

For turn six, Student C reverted to AC. Although the words in this prompt show much discontinuity with previous prompts, the one-shot exemplar appears to extend ideas about books and school from prompts one, two, three and five:

*Right when we finished the book I realized I had to open my locker because I forgot to get out my books. And it was almost the live morning assembly.*

For turn seven, Student C wrote explicit instructions for the chatbot, "Continue the story," beginning with imperative verbs and not providing any additional input data or output exemplars (NLI). The instructions are the exact words found in prompts one, three and five, but the prompt is stripped of the one-shot exemplar and a question. In this way, this is Student C's first prompt that makes use of the chatbot's chat memory.

For turn eight, Student C reverted to AC. This one-shot exemplar appears to be extend ideas about books and schools from previous one-shot exemplars as the prompt shows continuity in words with those previous prompts:

*I quickly grabbed my books from my locker and rushed to the assembly hall. I was running late and the assembly had already started. I quickly found a seat and tried to settle in*

*without drawing too much attention. Luckily, I managed to get through the assembly without anyone noticing my late arrival.*

For turn 9, Student C wrote natural language instruction with a one-shot exemplar (AC + NLI). The one-shot exemplar appears to extend ideas from previous one-shot exemplars in the pathway.

### 4.2.4. Student D's Prompt Engineering Pathway

Student D was writing a blog entry about her school day. Student D used the Sage, Dragonfly and ChatGPT chatbots.

Figure 6 is a network graph for Student C's prompt engineering pathway. Student D took 11 turns to complete her writing task and wrote four types of prompts.

**Figure 6**

*Network graph of Student D's prompt engineering pathway*

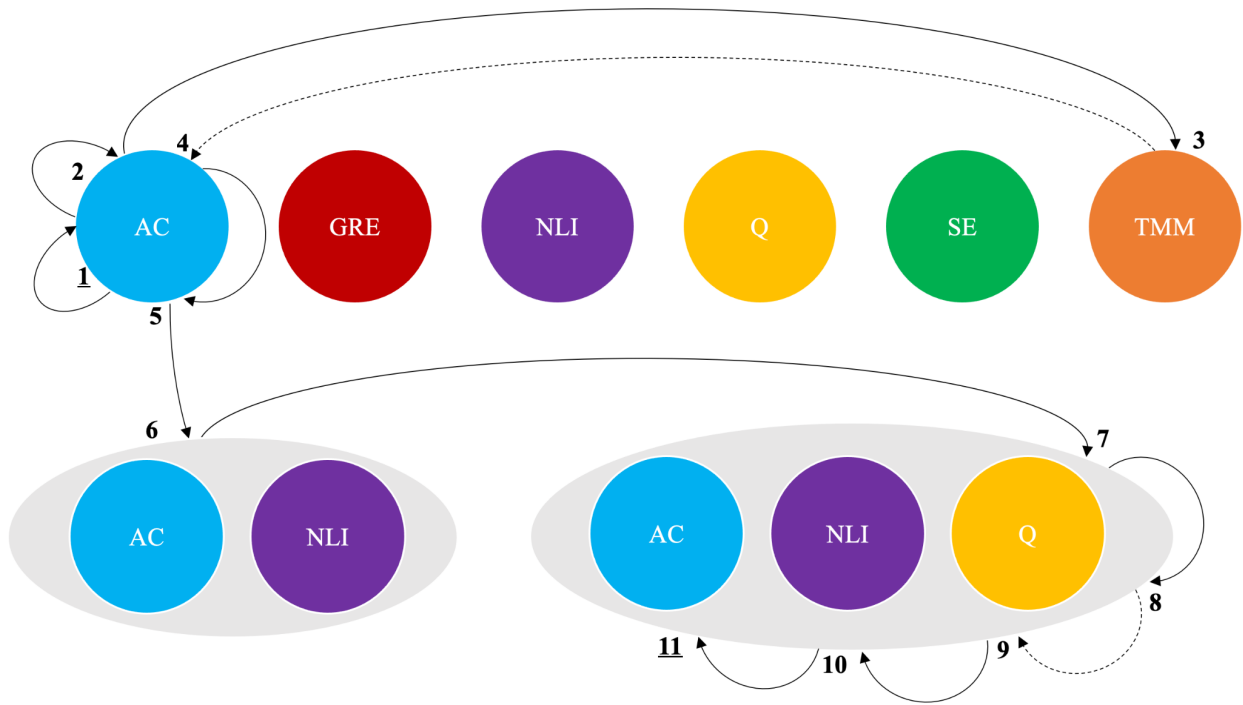

Student D began with the Sage chatbot. For turns one and two, Student D wrote prompts without natural language instruction (AC), intending for the chatbot to learn the input data exemplar and autocomplete the exemplar. The words in the prompt show much continuity, as Student D carefully and systematically changed words from prompt one to two:

- I was walking to the classroom at the first recess. (turn one)
- I was walking to the classroom. I walked there with Emily. I felt so happy. (turn two)

For turn three, Student D selected a prompt automatically suggested by the chatbot (TMM), "Tell me more." This appears to be the first and only time that Student D makes use of the chatbot's chat memory.

From turn four, Student D used the Dragonfly chatbot. For turns four and five, reverted to AC, using the exact same words for both prompts, "I was walking back to the classroom at the recess." The words in the prompts show much continuity with the AC prompts from turns one and two as Student D carefully and systematically changed a few words.

For turn six, Student D wrote natural language instruction with a one-shot exemplar (AC + NLI), "I was walking back to the classroom at the recess. Continue the story." The one-shot exemplar uses the exact words of prompts four and five.

For turns seven through nine, Student D wrote natural language instructions with one-shot exemplars and questions (AC + NLI + Q). These three prompts use the same words, "I was walking back to the classroom at the recess. Continue the story and answer this question : with whom did you do it with?" The one-shot exemplar in the three prompts follow the exact words of the one-shot exemplars from prompts four, five and six. The natural language instruction shows continuity with the words from prompt six and appears to be an extension of the natural language instruction.

>From turn nine, Student D used ChatGPT.

For turns 10 and 11, Student D wrote AC + NLI + Q prompts:

- I was breathing while waiting to go up to the classroom. Continue the story and answer this question : why did you do it? (turn 10)
- I am so happy that I walked back with Emily because I didn't like walking back to the classroom by myself. Continue the story and answer this question : how did you do it? (turn 11)

The natural language instruction in these prompts is the same as that found in prompts seven, eight and nine. On the other hand, Student B carefully and systematically changed the question words in prompts 10 and 11. In addition, Student B's one-shot exemplars in prompts 10 and 11 appear to be extensions of ideas already found in prompts one and two, and four through nine.

## 5. Discussion

### 5.1. Summary

#### *5.1.1. Types of Prompt Content*

Although the study has identified six distinct types of prompt content, no student had prompted with all six types. We interpret this as a matter of ignorance because students had not received any instruction on different types of prompt content and may not have known of different types of prompt content.

Of the six distinct prompt types that students' had used through a trial-and-error process, we found some prompt types such as AC and NLI are recommended in the literature (Reynolds

& McDonell, 2021) for effective prompting. On the other hand, we did not observe other types of prompts such as few-shot prompts, COT prompts (Kojima et al., 2022) and prompts for a chatbot's specific role or perspective (Akin, 2023). These have been recommended in the literature for effective prompting and we interpret their absence from pathways as either a matter of ignorance or related to the specific writing task.

Although students are completing an EFL writing task, we did not observe students prompt chatbots for explicit English language support, for example, proofreading and suggesting alternative phrasing. Students may not have prompted chatbots in this way because of ignorance, that is, their not knowing the SOTA chatbots' capabilities in these areas. Otherwise, these EFL students may not have needed this language support, because they felt sufficiently capable in English language writing. Ultimately, the absence of many prompt types in pathways suggests that students are unsophisticated users of chatbots, and are not optimizing their use of SOTA chatbots.

*5.1.2. Patterns of Prompt Engineering Pathways*

We observed Students A and B began with a type of prompt but never reverted to that prompt type in their pathway. In addition, all students used at least one prompt type only once in their pathway. We interpret these patterns as students exploring different prompt types and learning by trial-and-error whether a particular prompt type leads a chatbot to generate a desired output.

AC and AC + NLI + Q were the most popular prompt types. We interpret the AC prompt's use by Students A, C and D as an extension of students' existing experience using open-source text generators, which are less capable in performing human language tasks than SOTA chatbots like ChatGPT and were not trained to respond effectively to natural language instruction. In addition, we interpret the frequent use of AC and AC + NLI + Q prompts by Students C and D as these students learning to craft more sophisticated prompts, comprising different types of prompt content to which SOTA chatbots can respond effectively. In general, we view Students B, C and D's use of more sophisticated prompts as indicative of their learning to craft more sophisticated prompts.

We observed all students develop their respective ideas for their blog entry during their prompt engineering. In particular, we observed Students A, B and D develop ideas for their topic by making minor modifications to their words from previous prompts, which is a principle for effectively prompting a SOTA chatbot (Atlas, 2023). Alternatively, we observed Student C's prompts included extensions of ideas that had been used in previous prompts. We interpret these as students' exploring SOTA chatbots' capability to understand nuance in language and as students' refining their prompts so as to improve or to extend previous output. Similarly, Students C and D used more than one chatbot and repeated the same prompt with more than one chatbot. Furthermore, we observed that almost all repeated prompts between chatbots included an AC prompt type. We interpret these patterns as students' testing the output capabilities of each chatbot, particularly for idea generation with which EFL students have struggled (Crossley et al., 2016; Hayes & Flower, 2016).

## 5.2. Implications

This study reported cases of EFL students using ChatGPT and other SOTA chatbots for the first time to complete a writing task. Our cases illustrate these non-technical users' learning to engineer prompts through individual trial-and-error (Dang et al., 2022). The cases evidence that different prompt engineering pathways for the same writing task can emerge from students' trial-and-error process. At the same time, some pathways appeared more sophisticated than others.

Practically, the findings highlight the role of education and time in developing prompt engineering skills. In the context of EFL classroom writing, students would benefit from prompt engineering education if they are to realize the capabilities of chatbots to support student writing. At a very basic level, prompt engineering education may introduce students to chatbots as a type of AI, the capabilities of SOTA generative AI chatbots and different types of SOTA chatbots. Students can be introduced to prompting, and types of prompts that may unlock specific capabilities in chatbots. Furthermore, education may focus on which capabilities may be necessary to complete a particular writing task and the prompt types that unlock those capabilities.

To personalize prompt engineering education, the teacher's challenge is to identify a student's needs and to guide the student on an appropriate prompt pathway. For instance, an educator can identify a student's intended use of a chatbot output for the writing task and recommend a specific type of prompt. Besides, a student may also have their individual prompt engineering needs met by learning to engineer prompts alongside their classmates. By this individual and corporate education in the EFL writing classroom, students can become more sophisticated prompt engineers, realizing different prompt types and combinations of prompt types.

## 5.3. Limitations and Future Research

This study has some limitations that may open up avenues for future research. Firstly, the number of cases in this study was relatively small, consisting of only four. In this way, the number of cases contributes to theoretical generalizations for their context and the cases were not intended for statistical generalization. However, future studies should consider increasing the number of cases in this and other contexts to enhance the external validity of our study's theoretical generalizations or alternatively, to establish other theoretical generalizations.

Participants in this study were asked to perform a timed, familiar writing task in class. It is possible that their prompting behaviors may differ in an untimed setting or when working on an unfamiliar writing task or a different text type. Future studies should explore these variables to obtain a more comprehensive understanding of students' prompt engineering. For example, longitudinal studies may be conducted to examine students' growth in prompt engineering with the support of their teachers and peers. Similarly, while this study identified differences in prompt engineering pathways among the participants, it is necessary to investigate individual variables that may affect students' prompt engineering, such as gender, age, digital literacy, and writing skills. Future studies could explore these individual differences to provide a more nuanced understanding of students' prompt engineering.

This study focused on students' machine-in-the-loop writing at the stage of crafting prompts. Future research could expand the scope of machine-in-the-loop study and investigate students' evaluation of chatbots' output and integration of such output into their writing. These cases would provide a more comprehensive picture of human-AI collaborative writing. Likewise, it would be interesting to examine the quality of students' writing products and explore any associations with their prompt engineering pathways.

# 6. References

Akın, F. K. (2023). *The Art of ChatGPT Prompting: A Guide to Crafting Clear and Effective Prompts*. Gumroad. https://app.gumroad.com/d/1dd959d6c1edf9c4b4f5757ebe231f49

Atlas, S. (2023). *ChatGPT for Higher Education and Professional Development: A Guide to Conversational AI*. the College of Business at DigitalCommons@URI. https://digitalcommons.uri.edu/cba_facpubs/548

Author. (2022).

Author. (2023a).

Author. (2023b).

Bibauw, S., Van den Noortgate, W., François, T., & Desmet, P. (2022). Dialogue systems for language learning: A meta-analysis. *Language Learning & Technology*, *26*(1), 1–24. https://hdl.handle.net/10125/73488

Brown, T. B., Mann, B., Ryder, N., Subbiah, M., Kaplan, J., Dhariwal, P., Neelakantan, A., Shyam, P., Sastry, G., Askell, A., Agarwal, S., Herbert-Voss, A., Krueger, G., Henighan, T., Child, R., Ramesh, A., Ziegler, D. M., Wu, J., Winter, C., … Amodei, D. (2020). *Language Models are Few-Shot Learners* (arXiv:2005.14165). arXiv. https://doi.org/10.48550/arXiv.2005.14165

Caldarini, G., Jaf, S.F., & McGarry, K.J. (2022). A Literature Survey of Recent Advances in Chatbots. MDPI Journal Information*, 13*, 41. https://doi.org/10.3390/info13010041

Campbell, J. L., Quincy, C., Osserman, J., & Pedersen, O. K. (2013). Coding In-depth Semistructured Interviews: Problems of Unitization and Intercoder Reliability and Agreement. *Sociological Methods & Research*, *42*(3), 294–320. https://doi.org/10.1177/0049124113500475

Clark, E., Ross, A. S., Tan, C., Ji, Y., & Smith, N. A. (2018). Creative writing with a machine in the loop: Case studies on slogans and stories. *23rd International Conference on Intelligent User Interfaces*, Sydney, Australia, 329–340. https://doi.org/10.1145/3172944.3172983

Cohen, L., Manion, L., & Morrison, K. (2017). *Research Methods in Education*. Taylor & Francis Group.

Cohere. (2023). *Prompt Engineering*. Cohere AI. https://docs.cohere.com/docs/llmu

Creswell, J. W. (2013). *Research Design: Qualitative, Quantitative, and Mixed Methods Approaches* (Fourth). Sage Publications, Inc.

Crossley, S. A., Muldner, K., & McNamara, D. S. (2016). Idea generation in student writing: Computational assessments and links to successful writing. *Written Communication*, *33*(3), 328–354. https://doi.org/10.1177/0741088316650178

Dang, H., Mecke, L., Lehmann, F., Goller, S., & Buschek, D. (2022). *How to Prompt? Opportunities and Challenges of Zero- and Few-Shot Learning for Human-AI Interaction in Creative Applications of Generative Models*(arXiv:2209.01390). arXiv. http://arxiv.org/abs/2209.01390

De Wilde, V. (2023). Lexical characteristics of young L2 English learners' narrative writing at the start of formal instruction. *Journal of Second Language Writing*, *59,* 100960. https://doi.org/10.1016/j.jslw.2022.100960

DeCuir-Gunby, J. T., Marshall, P. L., & McCulloch, A. W. (2010). Developing and Using a Codebook for the Analysis of Interview Data: An Example from a Professional Development Research Project. *Field Methods*, *23*(2), 136–155. https://doi.org/10.1177/1525822X10388468

Elabdali, R. (2021). Are two heads really better than one? A meta-analysis of the L2 learning benefits of collaborative writing. *Journal of Second Language Writing*, *52*, 100788. https://doi.org/10.1016/j.jslw.2020.100788

Hayes, J. R., & Flower, L. S. (2016). Identifying the organization of writing processes. In Gregg, L. W., & Steinberg, E. R. (Eds.), *Cognitive processes in writing* (pp. 3–30). Routledge.

Joffe, H., & Yardley, L. (2004). Content and Thematic Analysis. In D. F. Marks & L. Yardley (Eds.), *Research Methods for Clinical and Health Psychology* (pp. 56–68). SAGE Publications.

Kim, H., Yang, H., Shin, D., & Lee, J. H. (2022). Design principles and architecture of a second language learning chatbot. *Language Learning & Technology*, *26*(1), 1–18. http://hdl.handle.net/10125/73463

Kojima, T., Gu, S. S., Reid, M., Matsuo, Y., & Iwasawa, Y. (2022). *Large Language Models are Zero-Shot Reasoners* (arXiv:2205.11916). arXiv. https://doi.org/10.48550/arXiv.2205.11916

Lampinen, A. K., & McClelland, J. L. (2018). *One-shot and few-shot learning of word embeddings* (arXiv:1710.10280). arXiv. http://arxiv.org/abs/1710.10280

Liu, P., Yuan, W., Fu, J., Jiang, Z., Hayashi, H., & Neubig, G. (2021). *Pre-train, Prompt, and Predict: A Systematic Survey of Prompting Methods in Natural Language Processing* (arXiv:2107.13586). arXiv. http://arxiv.org/abs/2107.13586

Long, D., & Magerko, B. (2020). What is AI Literacy? Competencies and Design Considerations. In *Proceedings of the 2020 CHI Conference on Human Factors in Computing Systems* (pp. 1–16). Association for Computing Machinery. https://doi.org/10.1145/3313831.3376727

Oppenlaender, J., Linder, R., & Silvennoinen, J. (2023). *Prompting AI Art: An Investigation into the Creative Skill of Prompt Engineering* (arXiv:2303.13534). arXiv. https://doi.org/10.48550/arXiv.2303.13534

Parnami, A., & Lee, M. (2022). *Learning from Few Examples: A Summary of Approaches to Few-Shot Learning* (arXiv:2203.04291). arXiv. http://arxiv.org/abs/2203.04291

Saldana, J. (2012). *The Coding Manual for Qualitative Researchers*. SAGE Publications. https://books.google.com.hk/books?id=V3tTG4jvgFkC

Saravia, E. (2022). Prompt Engineering Guide [Jupyter Notebook]. In https://github.com/dair-ai/Prompt-Engineering-Guide (Original work published 2022)

Su, Y., Lin, Y., & Lai, C. (2023). Collaborating with ChatGPT in argumentative writing classrooms. *Assessing Writing*, *57*, 100752. https://doi.org/10.1016/j.asw.2023.100752

Sun, T., & Wang, C. (2020). College students' writing self-efficacy and writing self-regulated learning strategies in learning English as a foreign language. *System*, *90*, 102221. https://doi.org/10.1016/j.system.2020.102221

Vasylets, O., & Marín, J. (2021). The effects of working memory and L2 proficiency on L2 writing. *Journal of Second Language Writing*, *52*, 100786. https://doi.org/10.1016/j.jslw.2020.100786

Vuorikari, R., Kluzer, S., & Punie, Y. (2022). *DigComp 2.2: The Digital Competence Framework for Citizens - With new examples of knowledge, skills and attitudes*. https://doi.org/10.2760/115376

Wang, Y., Kordi, Y., Mishra, S., Liu, A., Smith, N. A., Khashabi, D., & Hajishirzi, H. (2023). *Self-Instruct: Aligning Language Models with Self-Generated Instructions* (arXiv:2212.10560). arXiv. http://arxiv.org/abs/2212.10560

White, J., Fu, Q., Hays, S., Sandborn, M., Olea, C., Gilbert, H., Elnashar, A., Spencer-Smith, J., & Schmidt, D. C. (2023). *A Prompt Pattern Catalog to Enhance Prompt Engineering with ChatGPT* (arXiv:2302.11382). arXiv. http://arxiv.org/abs/2302.11382

Yan, D. (2023). Impact of ChatGPT on learners in a L2 writing practicum: An exploratory investigation. *Education and Information Technologies*. https://doi.org/10.1007/s10639-023-11742-4

Yao, S., Zhao, J., Yu, D., Du, N., Shafran, I., Narasimhan, K., & Cao, Y. (2023). *ReAct: Synergizing Reasoning and Acting in Language Models* (arXiv:2210.03629). arXiv. http://arxiv.org/abs/2210.03629

Yang, D., Zhou, Y., Zhang, Z., & Li, T. J.-J. (2022). AI as an active writer: Interaction strategies with generated text in human-AI collaborative fiction writing. *Joint Proceedings of the ACM IUI Workshops*, Helsinki, Finland, 10. CEUR-WS Team.

Yin, R. K. (2003). *Case study research: Design and methods* (3rd ed.). Sage Publications.

Zhang, R., Zou, D., & Cheng, G. (2023). Chatbot-based training on logical fallacy in EFL argumentative writing. *Innovation in Language Learning and Teaching*, 1–14. https://doi.org/10.1080/17501229.2023.2197417

Zhou, Y., Muresanu, A. I., Han, Z., Paster, K., Pitis, S., Chan, H., & Ba, J. (2023). *Large Language Models Are Human-Level Prompt Engineers* (arXiv:2211.01910). arXiv. http://arxiv.org/abs/2211.01910

Zotzmann, K., & Sheldrake, R. (2021). Postgraduate students' beliefs about and confidence for academic writing in the field of applied linguistics. *Journal of Second Language Writing*, *52*, 100810. https://doi.org/10.1016/j.jslw.2021.100810

**Appendix 1**

*Writing Task*

In September 2022, you wrote a blog entry about your day at school. **In April 2023, write another blog entry using your own words and AI words. You can use words from your original blog entry.** Highlight your words in red, and AI words in black.

1. Write at least two paragraphs:
2. In each paragraph, answer questions with your sentences.
    a. What did you do this morning?
    b. Where did you do it?
    c. When did you do it?
    d. With whom did you do it?
    e. Why did you do it?
    f. How did you do it?
    g. How did you feel?

**Appendix 2**

*Students' completed writing tasks*

STUDENT A

Hi,I am (Student A's name) and today I am gonna tell all of you about my stay in Turkey.I woke up and Ifound myself on a new day.The sun was shining and the sky was a bright shade of blue.I just love the weather of Turkey.I came here with my homie Lellium.First,we decided to go out and grab some breakfast.A cup of turkish tea is always refreshing.Then we went to Topkapi palace.Its a cultural place with Ottoman historical artifacts.We even drank Ayran.Its a refreshing and healthy yogurt drink.

   The Ayran made me refreshed.I think the yogurt was very organic.Then we went to Cappadocia. Cappadocia is famous for its fairy chimneys and beautiful landscapes.We took a hot air balloon ride above Cappadocia and the view was amazing.Lilium got very tired.I felt so amazed by the beauty of that place.In the evening,we just took some pictures by the river side in Istanbul,the view was spectacular.I had an amazing day and felt me lose a lot of stress.Tomorrow we are going back.I will miss Turkey a lot.Bye Bye guys!

STUDENT B

My morning

For many people,morning are a time of new beginnings and fresh start.Whether you're an early riser or a night owl.The morning is a unique time of the day that can set your tone for the rest of your day.In my morning,I will eat my breakfast to get energy for the day.on my way heading to school,I'd always listen tomy favourite music to wake myself up around 7:20 am. Sometimes I meets Charlotte or Vanessa on the mini bus,I will sitand chat with them,I put that as a hints before the lucky day.BecauseSpending times with friends can be the most enjoyable and rewarding experience in life.Going to school is like a challenging experience everyone will have in their life,I enjoyed doing it,I feel like learning is fun,I like knowing more things everyday

STUDENT C

I read a book this morning at school with my friends. I told them to read this book, called "The Secret" with me because nobody seemed to have any interest in it other than my friends. We had a great time discussing the story and the characters. We all had different opinions on the plot and we enjoyed debating our views. At the end of the book, we all agreed that the protagonist had made the right decision. I felt very happy and enjoyed that my friends agreed to read this book with me.

Right when we were done discussing about the book, i realised I had to open my locker because I forgot to get out my books. And it was almost the live morning assembly. I quickly grabbed my books from my locker. My friend's were already running there while I was just opening my locker! Well, it's kind of my fault for telling them to read books with me in the first place. After I got my books from my locker I quickly rushed to the assembly hall. I was running late and the assembly had already started. I quickly found a seat and tried to settle in without drawing too much attention. Luckily, I managed to get through the assembly without anyone noticing my late arrival.

When the assembly ended, I quickly made my way back to the class. I was relieved that I had made it in time, but I was still a bit embarrassed that I had been late because the teacher was waiting already and getting ready for the first class for us. As I walked into the classroom, I heard a few of my classmates talking about the book we had read earlier that morning. I joined in the conversation and we discussed our favorite parts of the story. I thought to myself, "there's still 15 minutes left before class starts, why not join them?" So I did.

After discussing a bit more about the book, we started talking about our plans for the weekend. One of my friends suggested that we all go to the movies together. We all agreed that it sounded like a great idea and we started planning the details of our outing. We talked about which movie we wanted to see, what time we should meet, and who was going to drive. It was a great way to end the conversation before class started.

STUDENT D

I was walking to the classroom at the first recess.I walked there with Emily.As I made my way back to the classroom after the recess, I noticed that I had forgotten my notebook on the bench outside. I quickly turned around and headed back to retrieve it, hoping that nobody had taken it.I was so happy that I walked back with Emily because I didn't like walking back to the classroom by myself.I walked back to the classroom with my friend Emily, feeling grateful for her company. As we walked, we chatted about our weekend plans and the upcoming school dance. I had always felt a little self-conscious about walking back to the classroom by myself, especially when everyone else seemed to have a friend or two to walk with. But with Emily by my side, I felt more confident and at ease.

I was breathing while waiting to go up to the classroom.I breathed with Lelliam.As I stood in line waiting to go up to the classroom, I took a deep breath and let it out slowly. I repeated this a few times, feeling the tension in my body start to ease away.

Finally, it was my turn to go up to the classroom. I took one more deep breath and headed inside, feeling more centered and ready to face the day.When we were walking up to the classroom, we were still breathing.

# Appendix 3

*Students' prompts for chatbots during the writing task*

| Student | Chatbot used | Prompt no. | Prompt content | Prompt type | Note |
|---|---|---|---|---|---|
| A | Sage | 1 | I woke up and | AC | |
| | | 2 | Some famous places in turkey | SE | |
| | | 3 | Famous drinks in Turkey | SE | |
| | | 4 | Some fun activities to do in turkey on vacation | SE | |
| | | 5 | Tell me more. | TMM | |
| B | Sage | 1 | hello | GRE | |
| | | 2 | Write a blog entry about a day going to secondary school | NLI | |
| | | 3 | Write a short blog entry about my day at school | NLI | |
| | | 4 | Write a blow entry about. My day to school | NLI | |
| | | 5 | What did you do this morning | Q | |
| | | 6 | Ideas to do in the morning before school | SE | |
| | | 7 | Write a blog entry about my morning | NLI | |
| | | 8 | Write a blog entry about morning | NLI | |
| | | 9 | What are some other ways to make mornings more enjoyable? | TMM + Q | Accidentally prompted |
| | | 10 | Write a blow entry about meeting friends is happy | NLI | |
| | | 11 | Write a blow entry about why going to school is a good thing and how it feels | NLI + Q | |
| | | | like | | |
| | | 12 | Why learning is fun | Q | |
| C | Sage | 1 | I read a book this morning at school. Continue the story and answer this question : with whom did you do it with? | AC + NLI + Q | |
| | | 2 | I read a book this morning at school.. | AC | |
| | ChatGPT | 3 | I read a book this morning at school. Continue the story and answer this question : with whom did you do it with? | AC + NLI + Q | |
| | Sage | 4 | Sage hi | GRE | |
| | Dragonfly | 5 | I read a book this morning at school. Continue the story and answer this question : with whom did you do it with? | AC + NLI + Q | |
| | | 6 | Right when we finished the book I realised I had to open my locker because I forgot to get out my books. And it was almost the live morning assembly. | AC | |
| | | 7 | Continue the story | NLI | |
| | | 8 | I quickly grabbed my books from my locker and rushed to the assembly hall. I was running late and the assembly had already started. I quickly found a seat and tried to settle in without drawing too much attention. Luckily, I managed to get through the assembly without anyone noticing my late arrival. | AC | |
| | | 9 | "When the assembly ended, I quickly made my way to my first class. l was relieved that I had made it in time, but l was still a bit embarrassed that I had been late. As I walked into the classroom, I heard a few of my classmates talking about the book we had read earlier that morning. I joined in the conversation and we discussed our favorite parts of the story. lt was a great way to start the day." Continue the story. | AC + NLI | |
| | | 10 | Tell me more. | TMM | Accidentally prompted |
| D | Sage | 1 | I was walking to the classroom at the first recess. | AC | |
| | | 2 | I was walking to the classroom. I walked there with Emily. I felt so happy. | AC | |
| | | 3 | Tell me more. | TMM | |
| | Dragonfly | 4 | I was walking back to the classroom at the recess. | AC | |
| | | 5 | I was walking back to the classroom at the recess. | AC | |
| | | 6 | I was walking back to the classroom at the recess. Continue the story. | AC + NLI | |
| | | 7 | I was walking back to the classroom at the recess. Continue the story and answer this question : with whom did you do it with? | AC + NLI + Q | |
| | | 8 | I was walking back to the classroom at the recess. Continue the story and answer this question : with whom did you do it with? | AC + NLI + Q | |
| | ChatGPT | 9 | I was walking back to the classroom at the recess. Continue the story and answer this question : with whom did you do it with? | AC + NLI + Q | |
| | | 10 | I was breathing while waiting to go up to the classroom. Continue the story and answer this question : why did you do it? | AC + NLI + Q | |
| | | 11 | I am so happy that I walked back with Emily because I didn't like walking back to the classroom by myself. Continue the story and answer this question : how did you do it? | AC + NLI + Q | |